# An Evaluation of Calibrated and Uncalibrated High-Resolution RGB Data in Time Series Analysis for Coal Spoil Characterisation: A Comparative Study

Sureka Thiruchittampalam[1,2], Bikram Pratap Banerjee[3], Nancy F Glenn[4], Simit Raval[5*]

[1,5] School of Minerals and Energy Resources Engineering, University of New South Wales, Sydney, NSW, 2052, Australia

[2] Department of Earth Resources Engineering, University of Moratuwa, Moratuwa 01400, Sri Lanka

[3] School of Surveying and Built Environment, University of Southern Queensland, Toowoomba, Queensland, 4350, Australia

[4] Department of Geosciences, Boise State University, Boise, ID, USA

[1,2] s.thiruchittampalam@unsw.edu.au, [3] Bikram.Banerjee@unisq.edu.au, [4] nancyglenn@boisestate.edu, [5*] simit@unsw.edu.au

**Abstract**

Minor errors in the spoil deposition process, such as placing stronger materials with higher shear strength over weaker ones, can lead to potential dump failure. Irregular deposition and inadequate compaction complicate coal spoil behaviour, necessitating a robust methodology for temporal monitoring. This study explores using unmanned aerial vehicles (UAV) equipped with red-green-blue (RGB) sensors for efficient data acquisition. Despite their prevalence, raw UAV data exhibit temporal inconsistency, hindering accurate assessments of changes over time. This is attributed to radiometric errors in UAV-based sensing arising from factors such as sensor noise, atmospheric scattering and absorption, variations in sun parameters, and variable characteristics of the sensed object over time. To this end, the study introduces an empirical line calibration with invariant targets, for precise calibration across diverse scenes. Calibrated RGB data exhibit a substantial performance advantage, achieving a 90.7% overall accuracy for spoil pile classification using ensemble (subspace discriminant), representing a noteworthy 7% improvement compared to classifying uncalibrated data. The study highlights the critical role of data calibration in optimising UAV effectiveness for spatio-temporal mine dump monitoring. The developed calibration workflow proves robust and reliable across multiple dates. Consequently, these findings play a crucial role in informing and refining sustainable management practices within the domain of mine waste management.

**Funding** This work was supported by the Australian Coal Industry's Research Program (ACARP) [Project number C29048]

**Declaration of interests** The authors declare that they have no known competing financial interests or personal relationships that could have appeared to influence the work reported in this paper.

# An Evaluation of Calibrated and Uncalibrated High-Resolution RGB Data in Time Series Analysis for Coal Spoil Characterisation: A Comparative Study

Sureka Thiruchittampalam[1,2], Bikram Pratap Banerjee[3], Nancy F Glenn[4], Simit Raval[5*]

[1,5] School of Minerals and Energy Resources Engineering, University of New South Wales, Sydney, NSW, 2052, Australia

[2] Department of Earth Resources Engineering, University of Moratuwa, Moratuwa 01400, Sri Lanka

[3] School of Surveying and Built Environment, University of Southern Queensland, Toowoomba, Queensland, 4350, Australia

[4] Department of Geosciences, Boise State University, Boise, ID, USA

[1,2] s.thiruchittampalam@unsw.edu.au, [3] Bikram.Banerjee@unisq.edu.au, [4] nancyglenn@boisestate.edu, [5*] simit@unsw.edu.au

**Abstract**

Minor errors in the spoil deposition process, such as placing stronger materials with higher shear strength over weaker ones, can lead to potential dump failure. Irregular deposition and inadequate compaction complicate coal spoil behaviour, necessitating a robust methodology for temporal monitoring. This study explores using unmanned aerial vehicles (UAV) equipped with red-green-blue (RGB) sensors for efficient data acquisition. Despite their prevalence, raw UAV data exhibit temporal inconsistency, hindering accurate assessments of changes over time. This is attributed to radiometric errors in UAV-based sensing arising from factors such as sensor noise, atmospheric scattering and absorption, variations in sun parameters, and variable characteristics of the sensed object over time. To this end, the study introduces an empirical line calibration with invariant targets, for precise calibration across diverse scenes. Calibrated RGB data exhibit a substantial performance advantage, achieving a 90.7% overall accuracy for spoil pile classification using ensemble (subspace discriminant), representing a noteworthy 7% improvement compared to classifying uncalibrated data. The study highlights the critical role of data calibration in optimising UAV effectiveness for spatio-temporal mine dump monitoring. The developed calibration workflow proves robust and reliable across multiple dates. Consequently, these findings play a crucial role in informing and refining sustainable management practices within the domain of mine waste management.

## 1 Introduction

The extraction of coal and the creation of coal spoil dumps pose significant environmental and safety challenges, especially when proper management practices are neglected. The haphazard disposal of these spoil materials onto large dumps without treatment or consideration for future use intensifies the environmental and safety issues related to coal mining. Given the contemporary emphasis on global sustainable waste management, there is a concerted effort to utilise and repurpose these waste materials thoughtfully. Moreover, a thorough understanding of coal spoil, encompassing factors such as shear strength and lithology, is crucial for guiding effective land restoration efforts (Fityus et al., 2008; Zevgolis et al., 2021). The comprehension of coal spoil characteristics, through informed decision-making, addresses environmental and safety challenges. This aligns with the global waste management priority and aims to mitigate the harmful impacts of coal mining. It adheres to principles of the international council on mining and metals (ICMM), which are related to effective risk management strategies (principle 4), health and safety (principle 5), and continuous improvement in environmental performance (principle 6) to manage adverse impacts.

While recognising the importance of understanding the geotechnical characteristics, the irregular deposition during placement introduces complexities and uncertainties in the behaviour of dumps over time (Lottermoser, 2007; Masoudian et al., 2019). Moreover, the inadvertent minor errors in the deposition process, specifically the misplacement of materials in an incorrect order - such as disposing of stronger materials with higher shear strength atop weaker ones with lower shear strength- can potentially culminate in the dump failure. These uncertainties underscore the need for a robust methodology that not only characterises coal spoil but also allows for the effective monitoring of its deposition over time as a dump is being formed. Traditional field methods, while valuable, face limitations in capturing the temporal dynamics of spoil material. This limitation necessitates the exploration of alternatives to enhance the temporal monitoring of coal spoil dumps.

Unmanned aerial vehicles (UAV) equipped with red-green-blue (RGB) sensors have recently been identified as a potential solution to the challenges associated with spoil pile characterisation. Thiruchittampalam et al. (2023b) utilised RGB images to characterise spoil piles at the pixel level, employing spectral and textural descriptors for this purpose. Further studies by Thiruchittampalam et al. (2023a) & Thiruchittampalam et al. (2024) applied object-based characterisation to the same task. These studies collectively demonstrate the feasibility of using RGB images for spoil pile characterisation. However, they are limited in their scope as they focus solely on a single time point. Overall, there is a gap in the literature concerning the temporal analysis of spoil pile dynamics.

The UAVs used in the mining field mostly do not incorporate irradiance correction sensors. Given the dynamic nature of dump environment changes, it is impractical to layout reflectance targets. Furthermore, implementing any field procedure to correct this in the mining field is fraught with risk. The raw data values obtained from UAV imagery display temporal inconsistency, which significantly impedes the accurate assessment of changes over time (Furby and Campbell, 2001). A pivotal aspect in the development of methodologies for temporal monitoring is the capacity to compare images captured on different dates. To facilitate this, it is essential to calibrate the digital numbers (DN) from each scene to standardised reference values. However, addressing the temporal dynamics of spoil piles and developing methods for standardising image data across different time points necessitates further studies. Such advancements will augment the accuracy and reliability of spoil pile characterisation and monitoring over time. Therefore, exploration in this field is of paramount importance.

Radiometric errors in UAV-based sensing, stemming from sensor noise, atmospheric scattering and absorption, changes in sun parameters, and the variable characteristics of the sensed object (Nansen et al., 2023), necessitate correction during preprocessing to ensure data accuracy. Several methods exist for radiometric correction, including empirical line calibration (ELC) (Daniels et al., 2023) and using calibration reference panels (CRP), downwelling light sensors (DLS), and radiative transfer models (Xue et al., 2023), each accounting for variations in radiometric correction. Out of these, the ELC is widely used with UAV data due to its simplicity and effectiveness in generating accurate estimates of surface reflectance. A unique feature of ELC is its use of multiple radiometric reference targets imaged at mission altitude, enhancing the calibration's precision. Moreover, ELC facilitates the recovery of information from shadowed areas, a frequent occurrence in high-resolution UAV imagery. This capability underscores the method's practicality in diverse imaging conditions (Poncet et al., 2019). Thus, the ELC stands as a robust, simple and reliable technique for radiometric correction in UAV-based sensing.

This study endeavours to contribute to the advancement of methodologies for temporal change monitoring in coal spoil environment. It seeks to elucidate the distinctions between uncalibrated and calibrated data-best performing classification algorithm pairings and aims to offer insights into the extent to which calibration contributes to significant improvements in the characterisation process. To achieve this, the study adapts a relative calibration technique known as scene-to-scene calibration. This approach incorporates empirical line calibration with invariant targets (ELC-IT) to facilitate precise calibration across diverse scenes, thereby enhancing accuracy and reliability of the data. Through these advancements, this study paves the way for the establishment and implementation of sustainable coal spoil management practices.

## 2 Methodology

### 2.1 Methodological overview

The goal of this study is to compare uncalibrated and calibrated data-algorithm pairings and determine the extent to which calibration contributes to improvements in the characterisation process. To this end, a comparative analysis was conducted to discern the disparities in performance between the best-performing classification algorithms utilising calibrated and uncalibrated data. The comprehensive workflow, encapsulating the stages from data acquisition to the final comparative evaluation of classification performance, is illustrated in Fig. 1.

The figure illustrates the balanced distribution of coal spoil categories (category 1 (Cat 1) and category 2 (Cat 2)) at the study site, emphasising the dataset's representative nature. The study uses ground truth data collected over seven time points in four months, during which UAV images were systematically captured. The workflow encompasses the processing of UAV images captured on selected dates to generate RGB orthomosaic and digital surface model (DSM), capturing the evolving nature of the spoil dump. ELC-IT is applied to obtain calibrated RGB data. Subsequently, object-based classification is conducted on both calibrated and uncalibrated data, involving Voronoi-based segmentation on DSM, feature extraction, and the integration of features and ground truth labels into machine learning pipelines. All the features derived over the seven time points were combined and used to train the machine learning algorithms. The final step involves the identification of the best-performing algorithms for calibrated and uncalibrated data, followed by a comprehensive performance comparison.

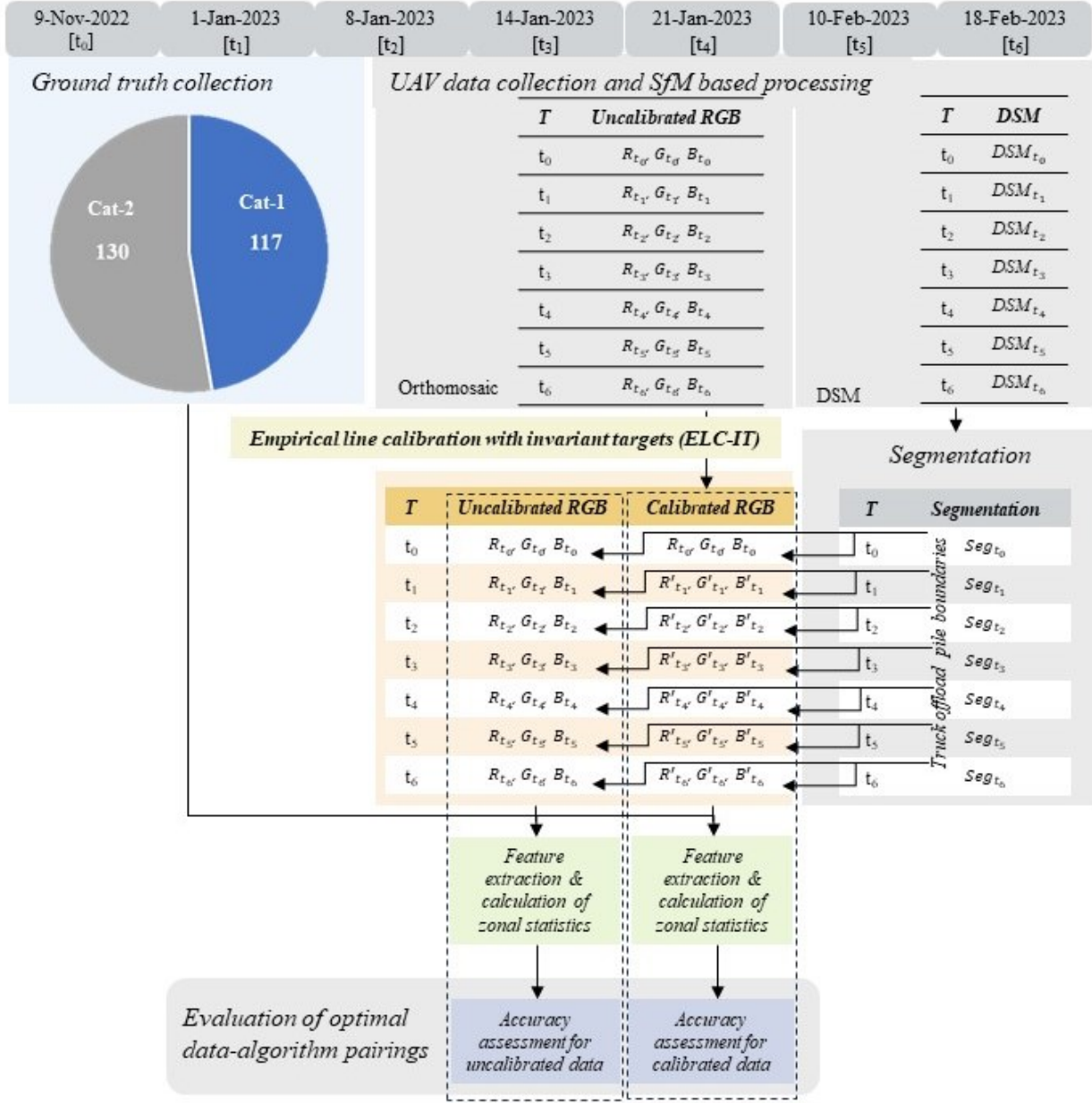

**Fig. 1.** Workflow for the comparative analysis of calibrated and uncalibrated data using the optimal algorithm pairing. Here, data is captured at seven time points, denoted as $t_0$, $t_1$, $t_2$, $t_3$, $t_4$, $t_5$, $t_6$. The Red, Green, Blue (R, G, B) bands captured during these seven time points are denoted as $R_{t_i \mid i=0,1,2,3,4,5,6}$, $G_{t_i \mid i=0,1,2,3,4,5,6}$, $B_{t_i \mid i=0,1,2,3,4,5,6}$ when uncalibrated. When calibrated, they are denoted as $R'_{t_i \mid i=1,2,3,4,5,6}$, $G'_{t_i \mid i=1,2,3,4,5,6}$, $B'_{t_i \mid i=1,2,3,4,5,6}$. Digital surface models (DSM) and segments of piles are denoted as $DSM_{t_i \mid i=0,1,2,3,4,5,6}$, $Seg_{t_i \mid i=0,1,2,3,4,5,6}$ respectively.

## 2.2 Study site and UAV data acquisition

The aerial imagery of the surveyed fields was acquired through the utilisation of a DJI M300 RTK drone (SZ DJI Technology Co., Ltd., Shenzhen, China). This UAV was equipped with a DJI Zenmuse P1 optical camera, thereby enabling the capture of high-resolution imagery essential for a detailed examination of spoil piles. The data acquisition activities transpired during seven time points, specifically on the 9th November 2022 ($t_0$), 1st January 2023 ($t_1$), 8th January 2023 ($t_2$), 14th January 2023 ($t_3$), 21st January 2023 ($t_4$), 10th February 2023 ($t_5$), and 18th February 2023 ($t_6$). Notably, the data collection sessions were scheduled around local solar noon to optimise lighting conditions.

To ensure accurate georeferencing of the acquired aerial imagery, five ground control points (GCP) were strategically positioned across the study area. The coordinates of these GCPs were measured using an Emlid Reach RS2 multi-frequency global navigation satellite system (GNSS) receiver. The receiver operated in kinematic survey mode and was connected to a networked transport of RTCM via internet protocol (NTRIP) correction service. This integrated setup guaranteed precise positioning of the GCPs, an imperative factor for attaining high-quality and accurately georeferenced aerial imagery.

The operational aspects of the UAV were programmed through the DJI Pilot app (SZ DJI Technology Co., Ltd). The UAV was operated at an altitude of 80 meters above ground level (AGL) to optimise imaging conditions for the Zenmuse P1 camera. This altitude was chosen to ensure comprehensive coverage, with an 80% forward and side overlap in imagery. Both cameras were positioned at nadir (90° with horizontal plane), and a singular grid flight path was established. The automatic capture mode was employed to facilitate the efficient acquisition of images. This methodical approach to survey planning and execution, along with the precise georeferencing methodology, contributed to the attainment of high-quality and accurate aerial imagery for subsequent analysis.

## 2.3 Conventional coal spoil characterisation approach and ground truth data collection

The foundation of coal spoil characterisation is rooted in the seminal study conducted by Simmons and McManus (2004). This investigation yielded the formulation of the spoil shear strength framework, illustrated in supplementary tables 1 and 2. This framework has since garnered widespread acceptance in Australia as a pivotal tool for classifying coal mine spoil. Its distinctive feature lies in its reliance on visual and tactile properties, circumventing the necessity for laborious laboratory tests to assess shear strength parameters.

In the present study, a comprehensive examination of 247 spoil piles within a dump area was undertaken. These spoil piles were systematically categorised into two groups: Cat-1 comprising 117 spoil piles and Cat-2 consisting of 130 spoil piles. The dump area itself comprised coal spoil piles arranged in a paddock-like configuration. Each spoil pile underwent characterisation according to the coal spoil characterisation framework, and their spatial coordinates were recorded utilising an Emlid Reach RS2 multi-frequency GNSS receiver.

The study employed a category-based stratified random sampling method to ensure that the spoil piles were evenly sampled. This method is useful in the spoil pile context when the spoil piles are heterogeneous and contain subgroups that differ significantly from each other. However, the accessibility of a spoil pile was a primary concern, and only spoil piles that were readily accessible were first included. Then, to increase the sample size, spoil piles exhibiting similar characteristics, discerned through visual observation and systematic analysis in the field, were ascribed identical attributes. This labelling procedure facilitated the creation of distinct categories of spoil piles sharing common characteristics, thereby enhancing the subsequent analysis and interpretation of the data. Fig. 2 depicts the orthomosaic generated over a seven time points, outlining the regions of paddock-tipped spoil piles that underwent the classification process and served as the locations for ground truth data collection.

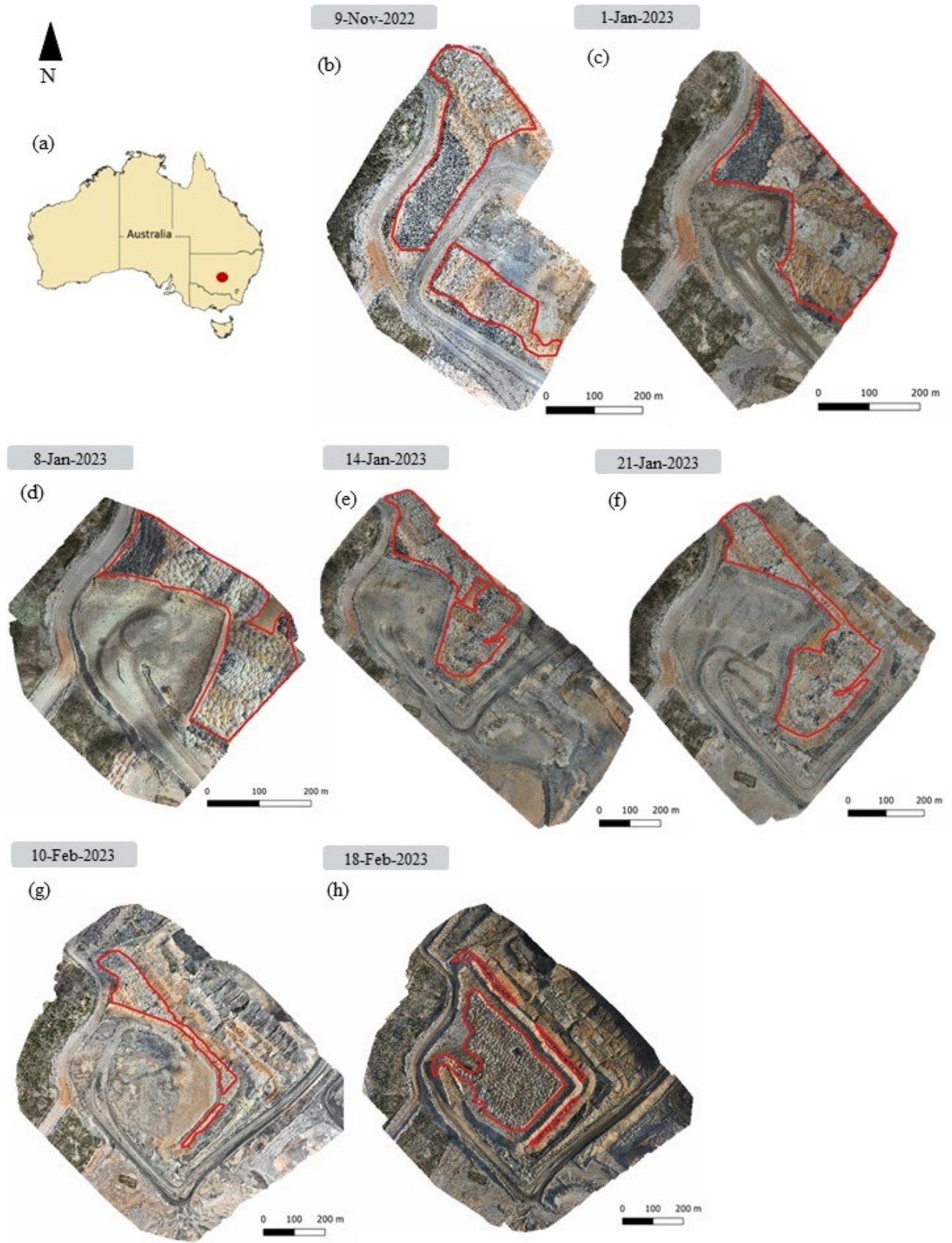

**Fig. 2. (a)** Geographic location of the selected mine site in New South Wales, Australia (exact coordinates withheld for proprietary reasons). The orthomosaic generated from the selected days exhibits discernible variations in digital numbers

within each scene. Regions of paddock-tipped spoil piles considered in this study on **(b)** 9$^{th}$ November 2022, **(c)** 1$^{st}$ January 2023, **(d)** 8$^{th}$ January 2023, **(e)** 14$^{th}$ January 2023, **(f)** 21$^{st}$ January 2023, **(g)** 10$^{th}$ February 2023, and **(h)** 18$^{th}$ February 2023 are demarcated.

## 2.4 UAV data processing

The acquired raw images from the UAV mission, encompassing optical images, underwent processing via Pix4Dmapper (Pix4D SA, Lausanne, Switzerland), a structure from motion (SfM)-based photogrammetric stitching package. The image sets were mosaicked by utilising captured images taken over the span of seven days, resulting in the generation of seven distinct orthomosaics (Fig. 2) and DSM.

The procedural workflow involved an alignment process aimed at identifying common features within overlapping images, facilitating the estimation of the camera's precise position and orientation for each image. This alignment process is imperative for the generation of accurate models. Subsequent to alignment, Pix4Dmapper executed camera calibration to ascertain both the intrinsic and extrinsic parameters of the camera. Intrinsic parameters were determined, encompassing focal length, lens distortion, and principal point, alongside extrinsic parameters denoting the camera's position and orientation within the scene.

GCPs were employed to refine the camera's positioning and orientation data, functioning as reference points to augment the accuracy of the resultant model. Pix4Dmapper utilised these GCPs to rectify errors in camera position and orientation, thereby facilitating the georeferencing of the resulting orthomosaic and DSM.

Following the processes of camera calibration and GCP integration, Pix4Dmapper produced the orthomosaic and DSM. The resultant orthomosaics exhibited ground sampling distances (GSD) of approximately 1.22 cm at an altitude of 80 m. Ultimately, seven distinct sets of orthomosaics and DSMs were generated, each characterised by a spatial resolution of 0.25 m and 0.50 m, respectively.

## 2.5 Temporal image calibration using empirical line calibration with invariant targets (ELC-IT)

The recorded raw data values obtained from the UAV sensor exhibit temporal inconsistency within the same geographical area, indicating that the sensor readings vary over time despite the location remaining constant.

In mine spoil pile regions, several factors contribute to data inconsistencies over time. The shadowing effect due to morphology of spoil piles, sun angle and illumination conditions significantly impacts the observed values, leading to data inconsistencies as the angle of the sun changes across time of year (de Souza et al., 2021). Atmospheric conditions, including temperature and humidity, also influence the observed values through atmospheric scattering and absorption processes. Sensor-related factors, such as sensor degradation or calibration drift, further contribute to the variability in the raw data values (Cao et al., 2019). Therefore, a comprehensive understanding of these influences is necessary for accurate interpretation and analysis of the raw data acquired by the UAV sensor system. This understanding can lead to more robust methods for data correction and analysis, thereby improving the accuracy of results in the study of mine spoil pile regions.

To address this challenge, a scene-to-scene correction methodology utilising invariant targets, ELC-IT, has been employed for the calibration of orthomosaics. Pseudoinvariant targets serve as key elements in establishing the relationship between successive pairs of images (Furby and Campbell, 2001). These targets, represented by image features such as invariant spoil piles in consecutive image pairs, are anticipated to exhibit constant reflectance characteristics over time. The temporal calibration process incorporates a robust ELC-IT procedure designed to estimate the association between the DN corresponding to invariant targets in two consecutive images. The presented calibration procedure is adaptable to sequences of images captured on different dates within the same geographical region. In each successive pair of images, 1000 random points within the overlapping area of unchanged spoil piles were selected for ELC-IT (Fig. 3). Specifically, the orthomosaic generated from UAV imagery captured on 9$^{th}$ November 2022, serves as the reference image for the subsequent orthomosaic generated on 1$^{st}$ January 2023. Subsequently, the calibrated orthomosaic from 1$^{st}$ January 2023, becomes the reference for calibrating the subsequent orthomosaic captured on 8$^{th}$ January 2023, and so forth (Fig. 4).

It is essential to underscore that this ELC-IT procedure can be applied to any imagery provided that atmospheric corrections exhibit linearity or be reasonably approximated by linear functions. The requirement for linearity in atmospheric corrections presupposes favourable atmospheric conditions. Consequently, image acquisition was scheduled for sunny days, precisely at local solar noon, with clear sky conditions, ensuring the requisite linearity for the atmospheric corrections. Additionally, to mitigate the potential impact of moisture on coal spoil, image capturing is deliberately avoided following rainfall. Areas with vegetation were intentionally excluded from the overlapping regions. This exclusion is due to the inherent variability of vegetation spectra, which are known to fluctuate significantly over short periods.

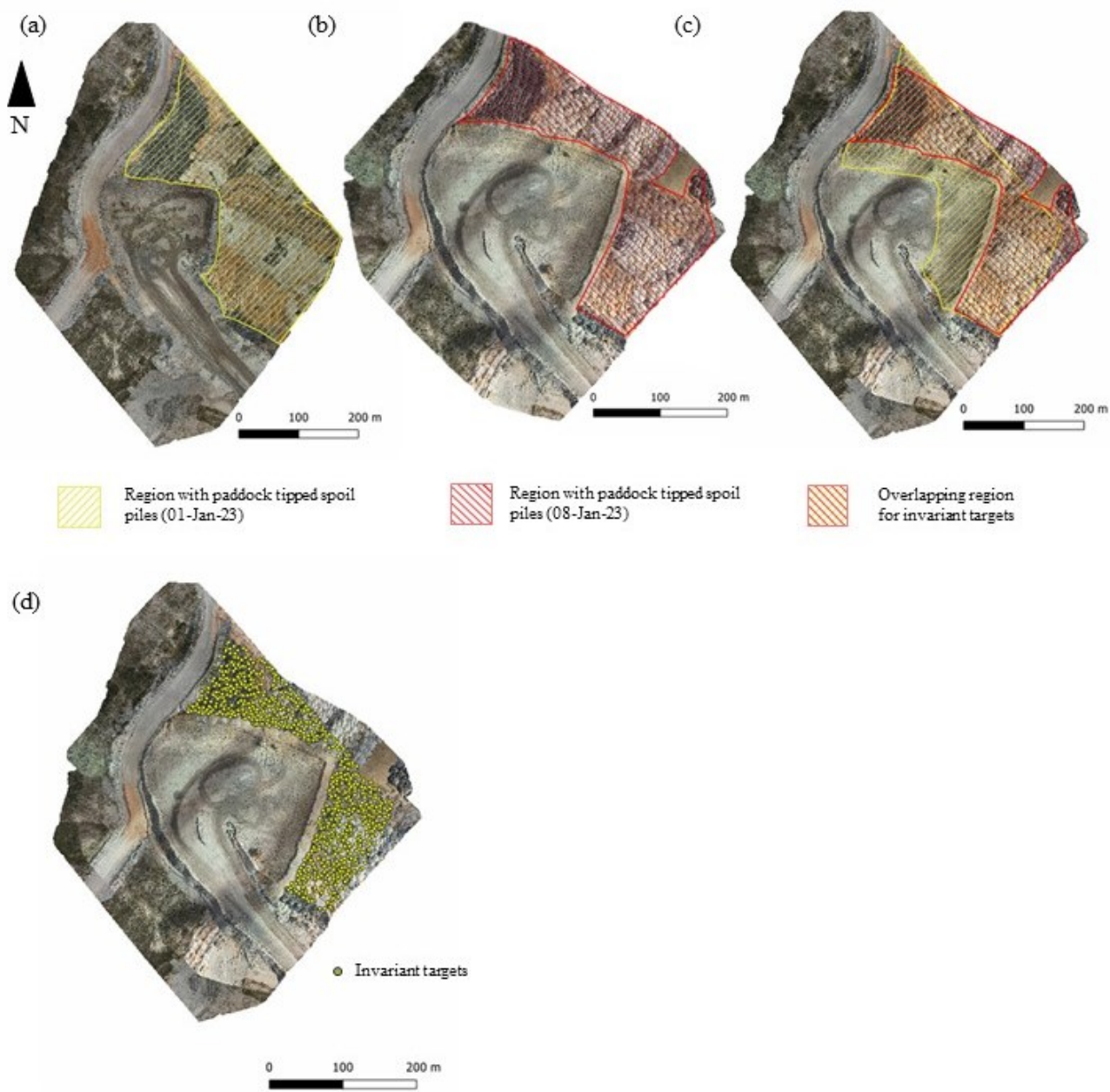

**Fig. 3.** Consecutive orthomosaic pairs captured on **(a)** 1$^{st}$ January 2023, and **(b)** 8$^{th}$ January 2023. **(c)** Overlapping regions in these orthomosaic pairs highlighting unchanged spoil piles. **(d)** Pseudoinvariant targets were generated within the overlapping unchanged region to facilitate empirical line calibration with invariant targets (ELC-IT).

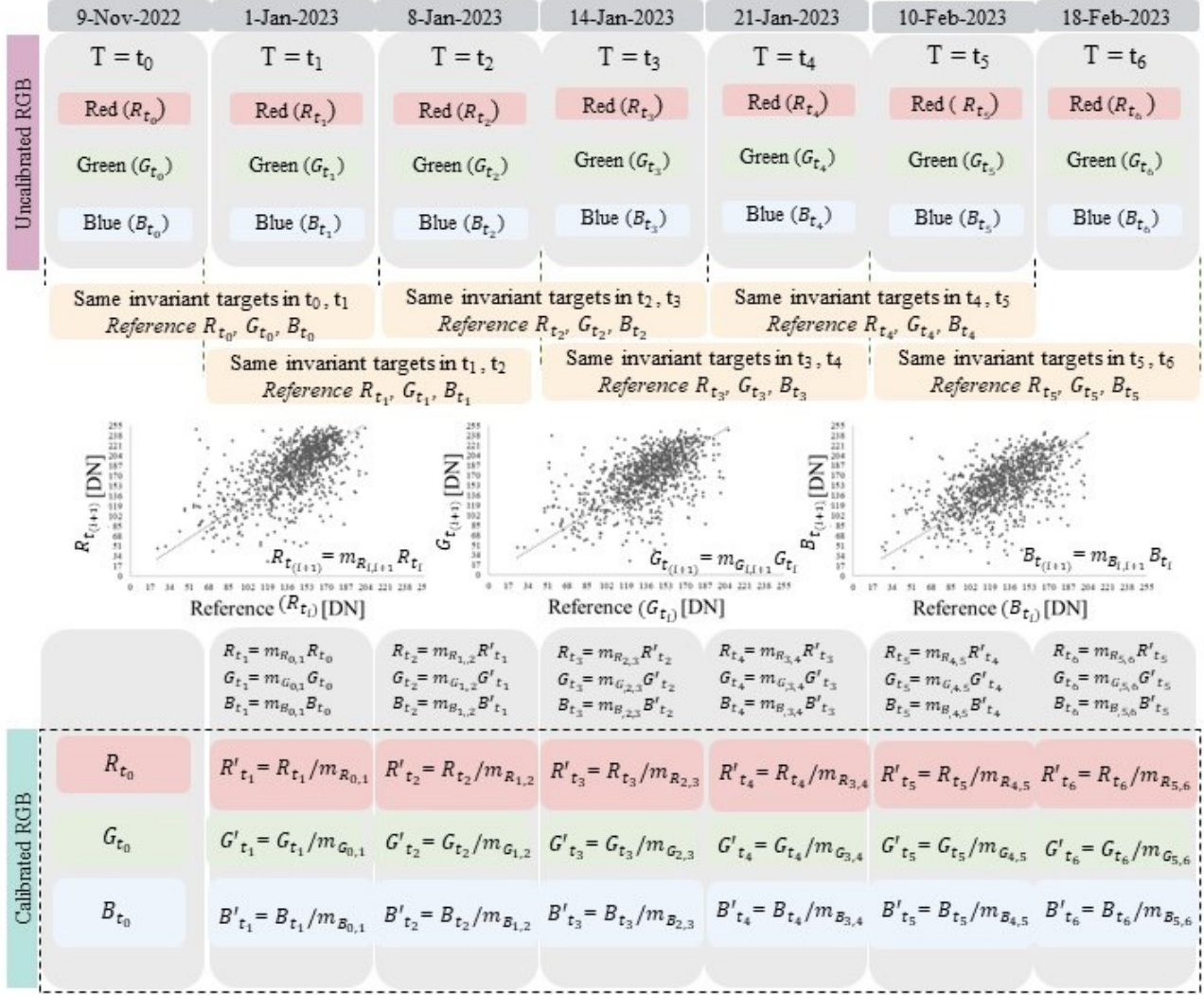

**Fig. 4.** Empirical line calibration with invariant targets (ELC-IT) procedure illustrating the estimation of the association between digital numbers (DN) of pseudoinvariant targets in two consecutive images and the subsequent calibration of seven orthomosaics.

## 2.6 Comparative analysis of calibrated and uncalibrated temporal data in object-based image analysis

The study utilised focused object-based classification for identification of heterogeneous spoil piles in mining areas. The procedural workflow encompassed within this study comprised several key stages, namely segmentation, feature extraction, the derivation of zonal statistics pertaining to features within the resultant segments and culminated in the ultimate process of classification.

The investigation employed morphology-based segmentation, Voronoi-based segmentation techniques, for the delineation of spoil piles, utilising DSMs as input data (Thiruchittampalam et al., 2023a). This segmentation technique, based on Voronoi tessellation, addressed irregular shapes through processes such as noise reduction, seed point detection, background seed point removal, and spoil pile polygonisation. Gaussian blurring with 12 sigma was employed to improve segmentation accuracy by smoothing out small irregularities in the DSMs (Nusantika et al., 2021). Local maxima detection and Otsu's thresholding method played crucial roles in seed point selection and distinguishing between background and spoil piles during segmentation.

Following the segmentation process, a comprehensive analysis of object characteristics and their interrelations was conducted through the utilisation of various features. This study categorises these features into four distinct groups: spectral, textural, structural (edge), and statistical features (Thiruchittampalam et al., 2024) (Table 1). Notably, the textural, structural, and statistical features were derived from individual bands, specifically the red, green, and blue bands of the orthomosaics. This approach was adopted to ensure the extraction of maximal information from the dataset. A total of 232 features were extracted by employing zonal statistics, specifically mean and standard deviation computations, within each Voronoi-based segment delineating each spoil pile. Feature extraction was executed on both uncalibrated and calibrated datasets for each of the seven orthomosaics individually.

**Table 1** Features extracted from red-green-blue (RGB) data

| Feature category | Features |
| --- | --- |
| Spectral | Red, Green, Blue, Red-Green ratio, Green-Blue ratio, Red-Blue ratio |
| Textural | Haralick features [Energy, Entropy, Correlation, Inverse Difference Moment, Inertia, Cluster Shade, Cluster Prominence, Haralick Correlation: kernel sizes - (3×3)]. Gabor: θ values of 0, π/4; σ values of 1,3; λ values of 0, π/4, π/2, 3π/4, and γ values of 0.05, 0.5. |
| Structural (Edge) | Sobel, Prewitt and Scharr use 3×3 kernel and Roberts edge operators use 2×2 kernel. Canny: Minimum and maximum values for double thresholding are chosen at 100 and 200, respectively. |
| Statistical | Gaussian: σ values of 3 and 7. Median: 3 × 3 kernel. |

Various machine learning algorithms, such as decision tree (Loussaief and Abdelkrim, 2018), discriminant analysis (Arabameri and Pourghasemi, 2019), naive Bayes (McCann and Lowe, 2012), support vector machine (SVM), k-nearest neighbors (kNN) (Loussaief and Abdelkrim, 2018), ensemble (Arboleda, 2019), neural network (Corenblit et al., 2023), and kernel approximation (Tien Bui et al., 2016), were employed for spoil pile classification. The study emphasised the necessity of multiple algorithms to assess accuracy in specific contexts, aiming to identify the most effective algorithm. Data partitioning, particularly the segregation of training and test sets, significantly influenced model performance (Lyons et al., 2018). A five-fold cross-validation approach ensured robust generalisation by iteratively designating one part for testing and employing the remaining four for training, with results averaged for an overall performance evaluation. The features obtained from both calibrated and uncalibrated datasets were individually subjected to five-fold cross-validation for comparison. The study employed overall accuracy and F-score metrics from confusion matrices to assess algorithmic performance, with F-score measuring the model's ability to confidently identify a class and overall accuracy evaluating the model's capacity to predict all classes.

## 3 Results

### 3.1 Temporal image calibration using empirical line calibration with invariant targets (ELC-IT)

The ELC-IT process of estimating calibration coefficients entails determining, for each image band, the optimal line of regression that best fits the data derived from both the reference image and the image intended for calibration. Linear regression analysis serves as a widely employed method for ascertaining the coefficients associated with the line of best fit. As illustrated in Fig. 5, a representative depiction of the calibration line, inclusive of the corresponding calibration equation, coefficient of determination ($R^2$), and p-value, is provided. The regression procedure is independently applied to each image band. The comprehensive calibration equation, $R^2$, and p-values for all image bands employed in the study are encapsulated in Table 2. The findings, as evidenced by the attainment of higher $R^2$ values in conjunction with statistically significant calibration equations at a confidence level of 95%, suggest the model's efficacy in capturing the inherent variability within the data.

The calibrated orthomosaics, derived through the application of these calibration equations, are showcased in Fig. 6. Notably, these calibrated orthomosaics exhibit consistent or closely aligned DNs across the scene, irrespective of acquisition dates, as compared to their uncalibrated counterparts. This consistency underscores the effectiveness of the calibration process in harmonising digital values and enhancing the interpretability and reliability of the orthomosaics.

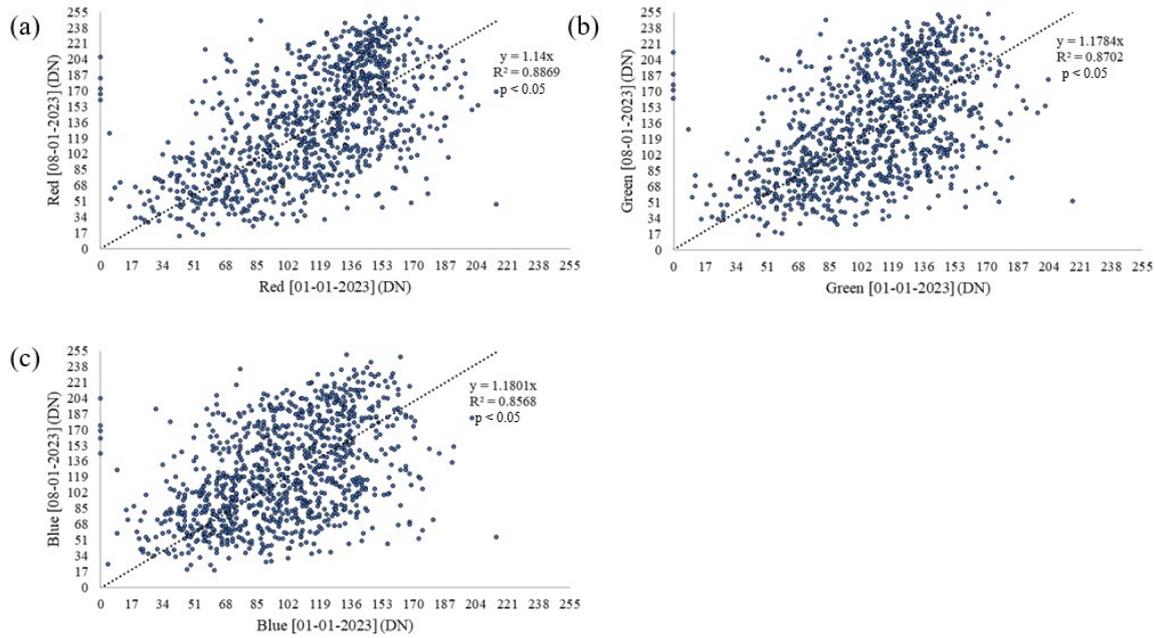

**Fig. 5.** Empirical line calibration with invariant targets (ELC-IT) depicting the relationship between digital numbers (DN) of invariant targets in reference image bands on 1st January 2023 and corresponding uncalibrated image bands on 8th January 2023 **(a)** red band, **(b)** green band and **(c)** blue band. The figure reports the coefficient of determination ($R^2$) and p-value as statistical indicators.

**Table 2.** Calibration equation derived through empirical line calibration with invariant targets (ELC-IT), featuring the coefficient of determination ($R^2$) which are statistically significant at a confidence level of 95%. The calibrated image bands were determined utilising invariant targets in reference image bands and corresponding uncalibrated image bands.

| Reference image bands | Uncalibrated image bands | Calibration equation | $R^2$ | Calibrated image bands |
|---|---|---|---|---|
| $R_{t_0}, G_{t_0}, B_{t_0}$ | $R_{t_1}, G_{t_1}, B_{t_1}$ | $R_{t_1} = 0.8987\ R_{t_0}$ | 0.8096 | $R'_{t_1} = R_{t_1} / 0.8987$ |
| | | $G_{t_1} = 0.8567\ G_{t_0}$ | 0.7917 | $G'_{t_1} = G_{t_1} / 0.8567$ |
| | | $B_{t_1} = 0.8172\ B_{t_0}$ | 0.764 | $B'_{t_1} = B_{t_1} / 0.8172$ |
| $R'_{t_1}, G'_{t_1}, B'_{t_1}$ | $R_{t_2}, G_{t_2}, B_{t_2}$ | $R_{t_2} = 1.14\ R'_{t_1}$ | 0.8869 | $R'_{t_2} = R_{t_2} / 1.14$ |
| | | $G_{t_2} = 1.1784\ G'_{t_1}$ | 0.8702 | $G'_{t_2} = G_{t_2} / 1.1784$ |
| | | $B_{t_2} = 1.1801\ B'_{t_1}$ | 0.8568 | $B'_{t_2} = B_{t_2} / 1.1801$ |
| $R'_{t_2}, G'_{t_2}, B'_{t_2}$ | $R_{t_3}, G_{t_3}, B_{t_3}$ | $R_{t_3} = 0.9257\ R'_{t_2}$ | 0.9 | $R'_{t_3} = R_{t_3} / 0.9257$ |
| | | $G_{t_3} = 0.9258\ G'_{t_2}$ | 0.892 | $G'_{t_3} = G_{t_3} / 0.9258$ |
| | | $B_{t_3} = 0.9048\ B'_{t_2}$ | 0.8861 | $B'_{t_3} = B_{t_3} / 0.9048$ |
| $R'_{t_3}, G'_{t_3}, B'_{t_3}$ | $R_{t_4}, G_{t_4}, B_{t_4}$ | $R_{t_4} = 0.983\ R'_{t_3}$ | 0.901 | $R'_{t_4} = R_{t_4} / 0.983$ |
| | | $G_{t_4} = 0.9769\ G'_{t_3}$ | 0.8968 | $G'_{t_4} = G_{t_4} / 0.9769$ |
| | | $B_{t_4} = 0.9432\ B'_{t_3}$ | 0.8887 | $B'_{t_4} = B_{t_4} / 0.9432$ |
| $R'_{t_4}, G'_{t_4}, B'_{t_4}$ | $R_{t_5}, G_{t_5}, B_{t_5}$ | $R_{t_5} = 1.2659\, R'_{t_4}$ | 0.9631 | $R'_{t_5} = R_{t_5} / 1.2659$ |
| | | $G_{t_5} = 1.2196\, G'_{t_4}$ | 0.9585 | $G'_{t_5} = G_{t_5} / 1.2196$ |
| | | $B_{t_5} = 1.1824\, B'_{t_4}$ | 0.952 | $B'_{t_5} = B_{t_5} / 1.1824$ |
| $R'_{t_5}, G'_{t_5}, B'_{t_5}$ | $R_{t_6}, G_{t_6}, B_{t_6}$ | $R_{t_6} = 0.5482\, R'_{t_5}$ | 0.7075 | $R'_{t_6} = R_{t_6} / 0.5482$ |
| | | $G_{t_6} = 0.5338\, G'_{t_5}$ | 0.7449 | $G'_{t_6} = G_{t_6} / 0.5338$ |
| | | $B_{t_6} = 0.5215\, B'_{t_5}$ | 0.7971 | $B'_{t_6} = B_{t_6} / 0.5215$ |

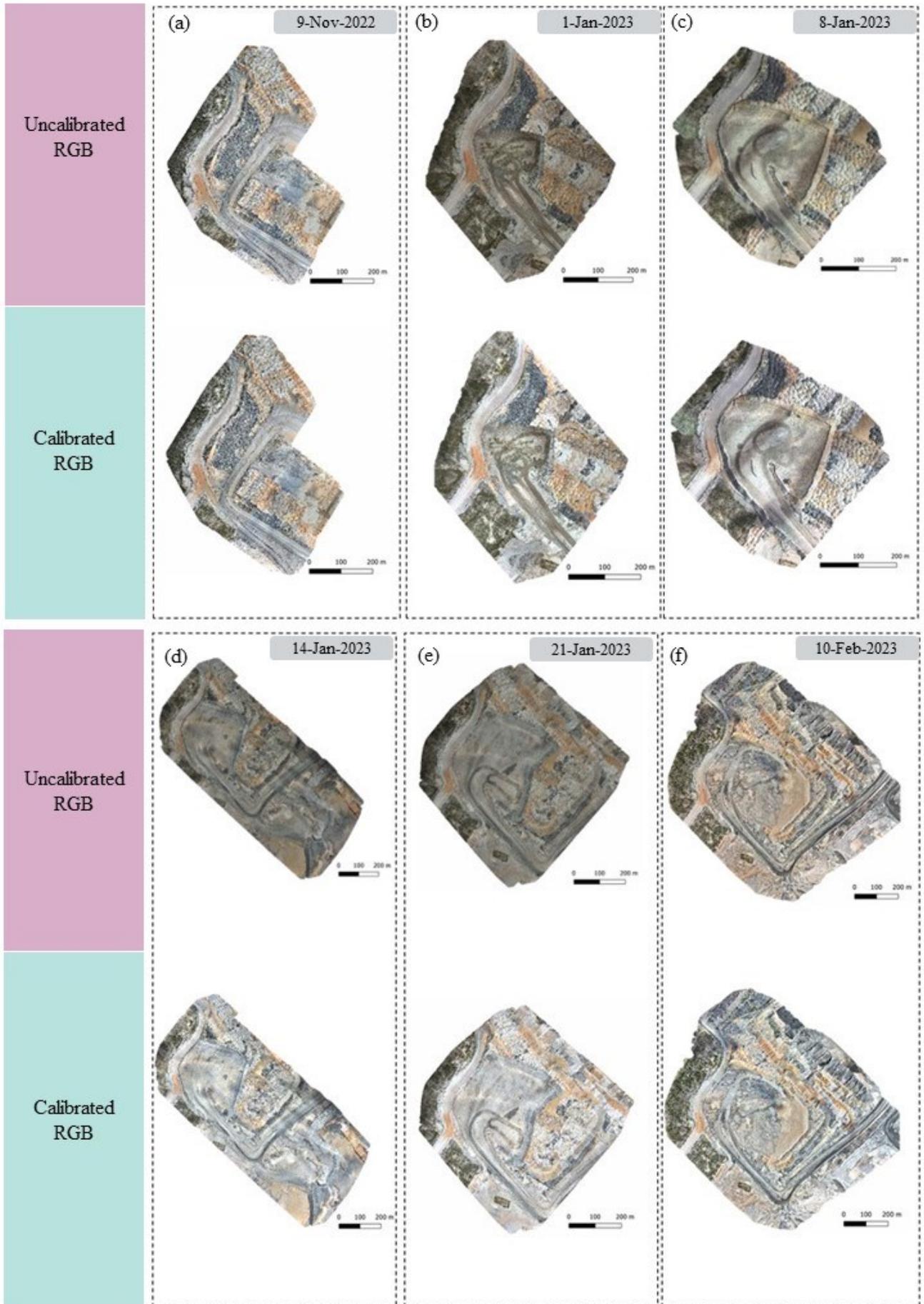

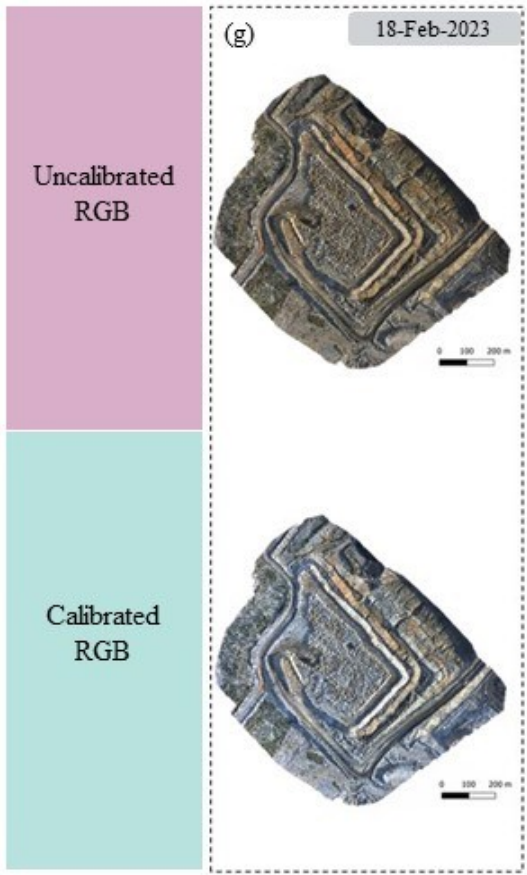

**Fig. 6.** Calibrated red-green-blue (RGB) orthomosaics and uncalibrated RGB orthomosaics on selected seven time points: **(a)** 9th November 2022, **(b)** 1st January 2023, **(c)** 8th January 2023, **(d)** 14th January 2023, **(e)** 21st January 2023, **(f)** 10th February 2023, and **(g)** 18th February 2023. Comparison of calibrated and RGB orthomosaics reveals consistent or closely matching digital numbers (DN).

### 3.2 Comparative analysis of calibrated and uncalibrated temporal data in object-based image analysis

Among the various algorithms assessed, the ensemble method, specifically subspace discriminant, exhibited superior performance when applied to calibrated RGB data, while the quadratic SVM demonstrated notable efficacy in handling uncalibrated RGB data. To ensure an equitable evaluation, the performance of both algorithms was systematically compared across the two types of data (Fig. 7(a)). In the context of calibrated data, the ensemble method (subspace discriminant) and quadratic SVM achieved accuracies of 90.7% and 85.8%, respectively. In contrast, when applied to uncalibrated data, the ensemble method yielded an accuracy of 50.6%, while the quadratic SVM demonstrated a higher accuracy of 83%. Notably, despite the quadratic SVM attaining the highest accuracy of 83% in uncalibrated data, the calibrated data still outperformed it with an accuracy of 85.8% when employing the same quadratic SVM model. The comparison of the best-performing algorithm-data pairings, specifically ensemble (subspace discriminant)-calibrated data and quadratic SVM-uncalibrated data, revealed an approximate 7% difference in accuracy. This disparity underscores the value added by the calibration process in enhancing the classification performance.

Further assessment of the best performing algorithm's efficacy is conducted through per-class accuracy metrics, specifically precision, recall, and F1-score, as depicted in Fig. 7(b) and (c). It is discerned that, for both categories, F-scores surpass 0.9 when features derived from calibrated data is employed. In contrast, features derived from uncalibrated data result in F-scores marginally exceeding 0.8 for both categories.

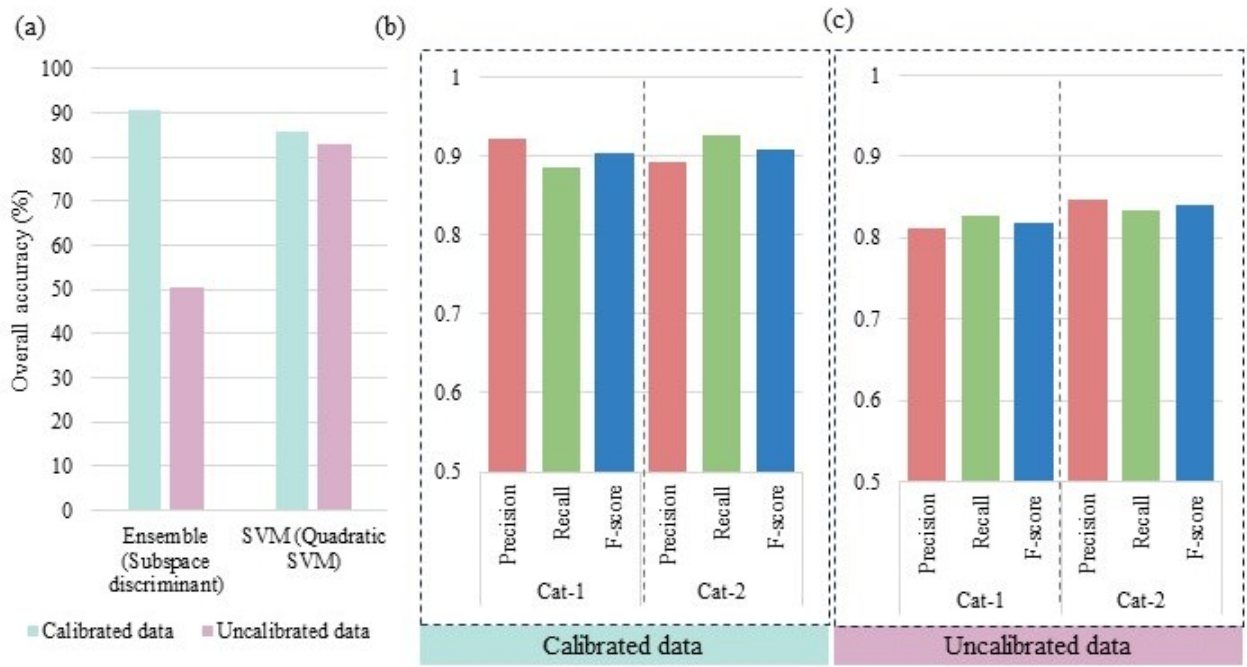

**Fig. 7. (a)** Comparative analysis of overall accuracy among the most effective machine learning algorithms applied to coal spoil categories, utilising input features derived from both uncalibrated and calibrated data. Precision, recall, and F-Score metrics for category classifications are presented for features obtained from **(b)** calibrated data and **(c)** uncalibrated data.

Fig. 8 showcase thematic maps generated by employing features derived from uncalibrated and calibrated RGB data as inputs for the most effective machine learning algorithms. These thematic maps elucidate the spoil categories and unveil variations in their distribution within the study area in response to the calibration process. The comparative analysis of predictions conducted on calibrated and uncalibrated datasets reveals that 73.8% of the predictions exhibit no discernible variance, while 26.2% of the prediction's manifest disparities. This suggests that, within a total area of 506860 m$^2$, discrepancies in predictions are observed in 132751 m$^2$.

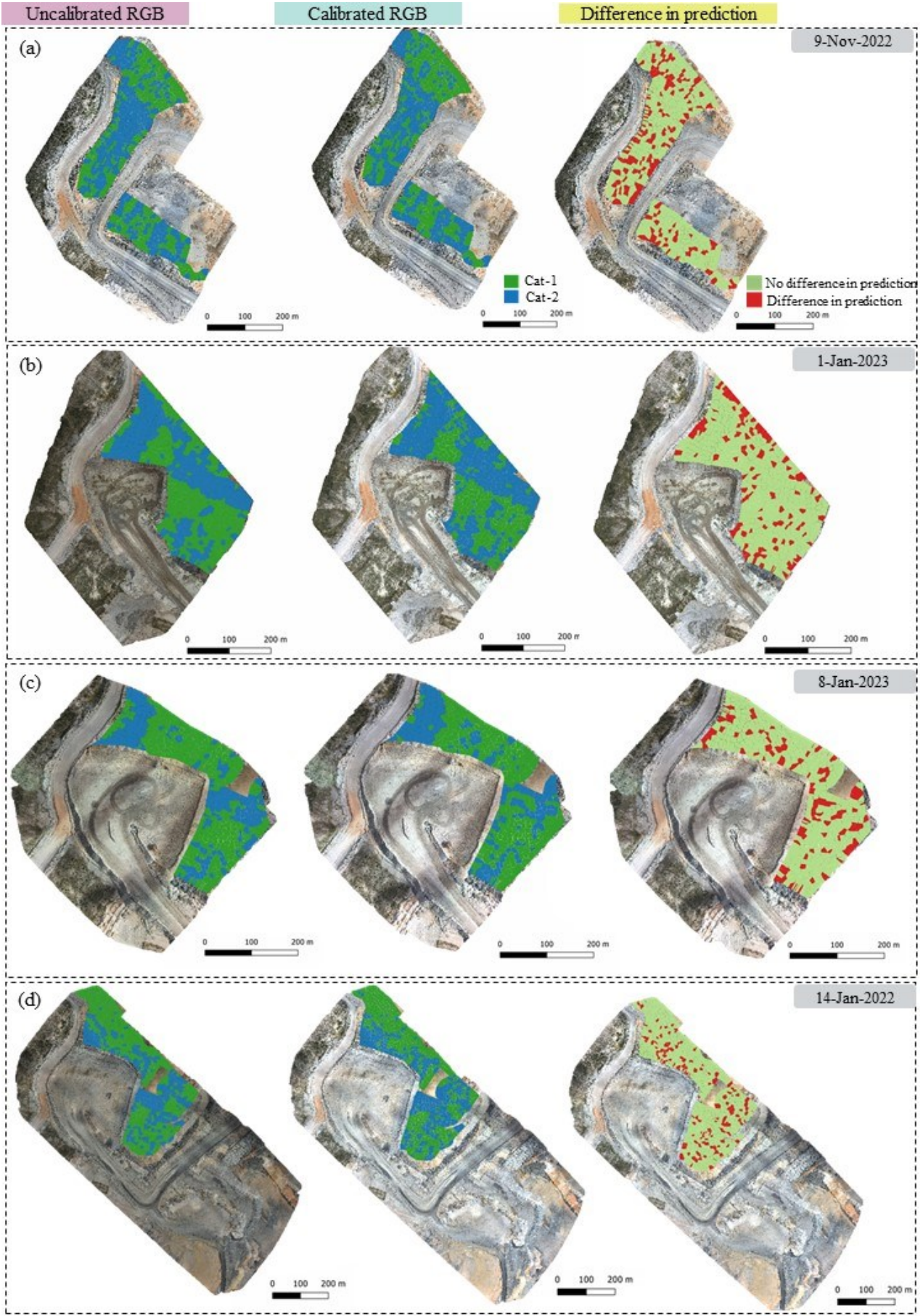

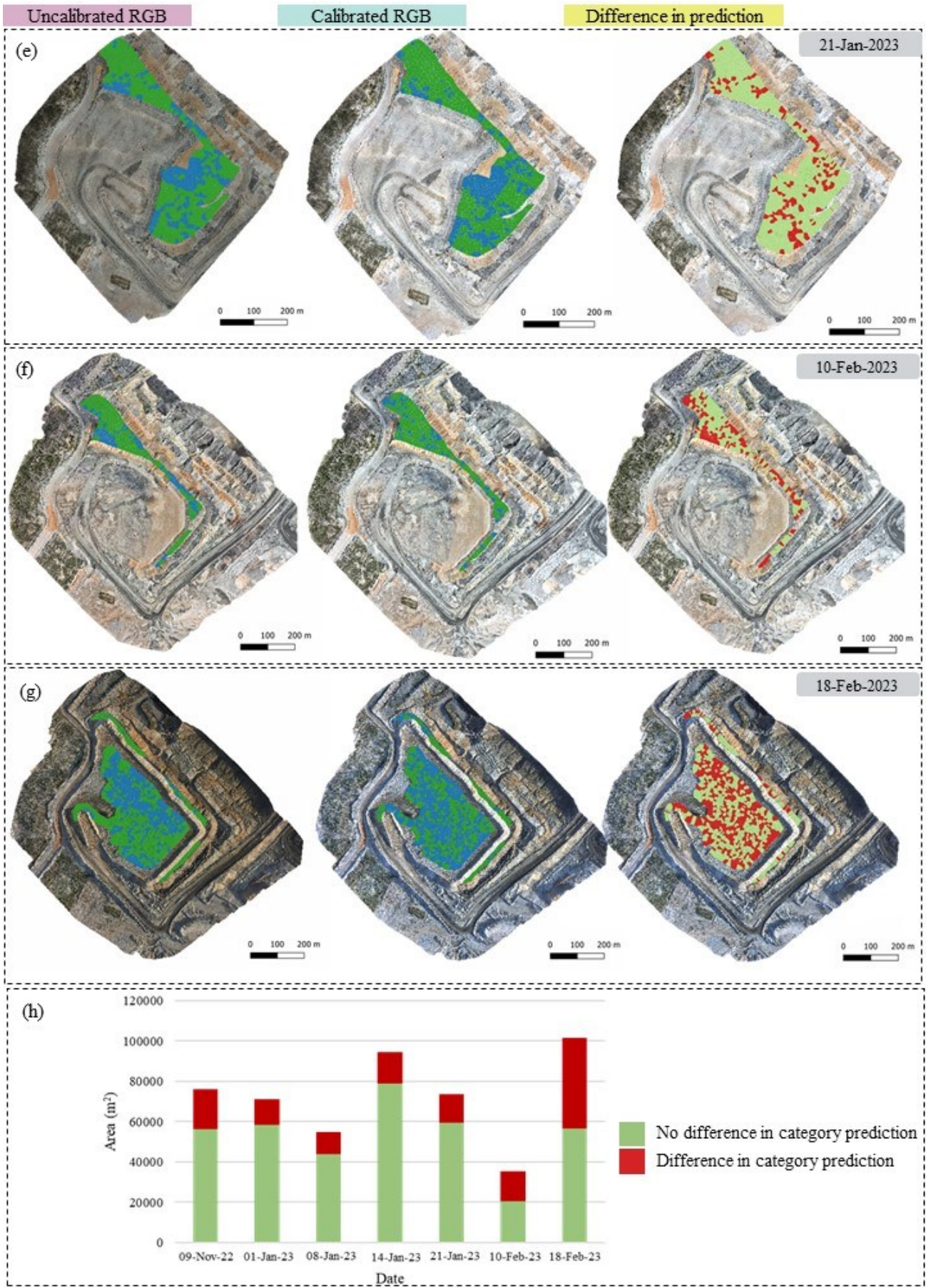

**Fig. 8.** Object-based classifications utilising features extracted from uncalibrated data through the quadratic support vector machine (SVM), calibrated data through ensemble (subspace discriminant), and the difference in predictions made on calibrated and uncalibrated data on specific dates: **(a)** 9th November 2022, **(b)** 1st January 2023, **(c)** 8th January 2023, **(d)** 14th January 2023, **(e)** 21st January 2023, **(f)** 10th February 2023, and **(g)** 18th February 2023. Additionally, **(h)** represents the area (in $m^2$) of predicted categories that exhibited variation across the seven specified time points.

## 4 Discussion

UAVs equipped with RGB sensors have become prevalent in mine environments, facilitating data acquisition; however, the recorded raw data values exhibit temporal inconsistency. An indispensable facet in the evolution of methodologies for temporal monitoring of changes in mine conditions is the capacity to compare images captured on different dates. This necessitates the calibration of DNs from each scene to standardised reference values. The calibration of images not only allows for the seamless display of image sequences with identical band combinations and shared enhancements but also expeditiously communicates alterations in the images. In instances where images remain uncalibrated, the collection of ground truths becomes imperative upon the acquisition of UAV images, entailing labour-intensive fieldwork. Conversely, calibrated images obviate the need for recurrent ground truth collection, as training data extracted once proves adequate for spatio-temporal classification across multiple images. The primary aim of this research is to elucidate the disparities inherent in spoil classification process when employing data-algorithm pairings that are either uncalibrated or calibrated. In addition, the study endeavours to offer insights into the degree to which calibration serves as a contributing factor to significant improvements in the spoil characterisation process within mine environments, specifically for the purpose of temporal analysis.

The evaluation of temporal consistency, accomplished through the application of ELC-IT to DNs derived from pseudo invariant targets, reference, and uncalibrated data, substantiates the resilience and dependability of the developed calibration workflow. It is imperative to emphasise that this calibration methodology is applicable to imagery, contingent upon the prerequisite that atmospheric corrections exhibit linearity or are reasonably approximated by linear functions. The necessity for linearity in atmospheric corrections presupposes favourable atmospheric conditions. Consequently, image acquisition was meticulously scheduled for sunny days, precisely at local solar noon, under clear sky conditions, thereby ensuring the requisite linearity for the atmospheric corrections (Furby and Campbell, 2001). The attainment of elevated $R^2$ values and diminutive p-values across all bands elucidates the statistical significance inherent in the ELC-IT equations. The resultant heightened confidence in the calibration process is reinforced by a 95% confidence level associated with the ELC-IT equations, thereby bolstering the overall reliability of the model. This discernible level of confidence assumes pivotal significance, as it ensures not only the statistical significance of the calibration process but also its adaptability across a spectrum of temporal scenarios. Furthermore, the validation of the model's robustness across multiple dates not only fortifies its immediate applicability but also extends its relevance for protracted monitoring and analysis endeavours. The demonstrated efficacy of the model over an array of temporal contexts enhances its standing as a potent tool for sustained and comprehensive environmental assessment, contributing to a nuanced understanding of temporal dynamics in the context of spoil characterisation.

The combination of Voronoi-based segmentation and machine learning classification in an integrated object-based approach (Thiruchittampalam et al., 2023a), particularly when applied to features from calibrated data, proves to be a robust method for classifying spoil piles. The calibrated RGB data demonstrates a notable performance advantage, attaining the highest overall accuracy of 90.7% for classification of spoil piles. This commendable achievement is realised through the strategic employment of an ensemble, specifically the subspace discriminant, as the classifier. Ensemble algorithms function by aggregating predictions from a collection of classifiers to categorise unclassified data. The subspace discriminant enhances the accuracy of its base learners (i.e., discriminants) through the implementation of the random subspace algorithm (Ashour et al., 2018). In contrast, the application of quadratic SVM to input features derived from uncalibrated RGB data yields a marginally lower overall accuracy of 83%. SVM is an algorithm that delineates optimal boundaries to separate classes by maximising the margin between them (Altay et al., 2020). Quadratic SVMs possess the capability to capture non-linear relationships inherent in the data. Unlike their linear counterparts, which establish linear decision boundaries, quadratic SVMs exhibit a heightened capacity to model intricate decision boundaries that align more closely with non-linear patterns present in the data.

The observed disparity of approximately 7% in accuracy between calibrated and uncalibrated data highlights the substantial impact of the calibration process on classification performance. The ensemble method's significant drop in accuracy from calibrated (90.7%) to uncalibrated data (50.6%) underscores the vulnerability of certain algorithms to data variations. In contrast, the quadratic SVM maintains robust performance in both scenarios, though calibrated data consistently outperforms uncalibrated data when employing the same model.

The study reveals that 73.8% of predictions showed no discernible difference, while 26.2% exhibited variations, translating to 132751m$^2$ of the surveyed area displaying disparities in prediction outcomes. These differences underscore the potential consequences of errors in the deposition process, such as the inadvertent misplacement of materials, which can lead to dump failure. The finding emphasises the value of incorporating a calibration process (ELC-IT), particularly in UAVs lacking irradiance sensors commonly found in mine sites, to enhance prediction accuracy and mitigate risks associated with deposition inaccuracies. However, ELC-IT works by measuring DNs of invariant targets in the scene. Hence, the calibration coefficients obtained are unique to the location containing the invariant targets. This means that the calibration is specific to the conditions at the time of image capture, which can limit the direct comparison of different locations over time. Continuous methodological enhancements, integrating irradiance sensors and improving the training dataset over the time, are recommended to improve the reliability of predictions and contribute to the stability and safety of dump formations in mining operations.

In summary, the study underscores the crucial role of data calibration in not only boosting overall accuracy but also improving per-class metrics and ensuring thematic map accuracy. Calibrating data before classification narrows the disparity between on-site and image-based characterisations, resulting in a noteworthy 7% increase in accuracy. Given the potential hazards associated with subtle inaccuracies in the deposition process leading to dump failures, the calibration procedure emerges as a pivotal step. This adaptation holds the potential to strengthen dump stability over an extended period, especially when dealing with a substantial volume of spoil deposition.

## 5 Conclusion

This study emphasises the critical role of data calibration in optimising the effectiveness of UAVs equipped with RGB sensors for mine environment monitoring. The findings reveal significant disparities in spoil classification processes between calibrated and uncalibrated data-algorithm pairings, highlighting the substantial improvements achievable through proper calibration. The developed calibration workflow proves robust and reliable, validated across multiple dates and demonstrating statistical significance. With 95% confidence, the calibration equations make the model reliable and adaptable across different time scenarios. The integration of Voronoi-based segmentation and machine learning classification, especially on features derived from calibrated data, emerges as a robust approach for spoil pile classification. Data calibration not only enhances overall accuracy but also improves per-class metrics. This approach minimises the gap between on-site characterisation and image-based characterisation, providing a potent tool for sustained and comprehensive environmental assessment. In essence, this study substantiates the transformative impact of data calibration on the precision and reliability of UAV-based monitoring in mine environments, contributing to a nuanced understanding of temporal dynamics in spoil characterisation.

## Author contributions

**Sureka Thiruchittampalam, Bikram Pratap Banerjee, Nancy F Glenn, Simit Raval:** Conceptualization; **Sureka Thiruchittampalam:** Data curation; **Sureka Thiruchittampalam:** Formal analysis; **Simit Raval:** Funding acquisition; **Sureka Thiruchittampalam:** Investigation; **Sureka Thiruchittampalam, Bikram Pratap Banerjee, Nancy F Glenn, Simit Raval:** Methodology; **Simit Raval:** Project administration; **Bikram Pratap Banerjee, Nancy F Glenn, Simit Raval:** Supervision; **Bikram Pratap Banerjee, Nancy F Glenn, Simit Raval:** Validation; **Sureka Thiruchittampalam:** Visualization; **Sureka Thiruchittampalam:** Writing - original draft; and **Bikram Pratap Banerjee, Nancy F Glenn, Simit Raval:** Writing - review & editing